\documentclass{article}

\usepackage{microtype}
\usepackage{graphicx}
\usepackage{subfigure}
\usepackage{booktabs} 

\usepackage{hyperref}

 \usepackage[accepted]{icml2023}

\usepackage{amsmath}
\usepackage{amssymb}
\usepackage{mathtools}
\usepackage{amsthm}

\usepackage[capitalize,noabbrev]{cleveref}

\theoremstyle{plain}
\newtheorem{theorem}{Theorem}[section]

\theoremstyle{definition}

\theoremstyle{remark}

\newtheorem{observation}{Observation}

\usepackage[textsize=tiny]{todonotes}

\icmltitlerunning{Controlling Type Confounding in Ad Hoc Teamwork with Instance-wise Teammate Feedback Rectification}

\begin{document}

\twocolumn[
\icmltitle{Controlling Type Confounding in Ad Hoc Teamwork with Instance-wise Teammate Feedback Rectification}




\begin{icmlauthorlist}
\icmlauthor{Dong Xing}{bmi,zju}
\icmlauthor{Pengjie Gu}{ntu}
\icmlauthor{Qian Zheng}{bmi,zju}
\icmlauthor{Xinrun Wang}{ntu}
\icmlauthor{Shanqi Liu}{zjuc}
\icmlauthor{Longtao Zheng}{ntu}
\icmlauthor{Bo An}{ntu}
\icmlauthor{Gang Pan}{bmi,zju}
\end{icmlauthorlist}

\icmlaffiliation{bmi}{The State Key Lab of Brain-Machine Intelligence, Zhejiang University, Hangzhou, China}
\icmlaffiliation{zju}{College of Computer Science and Technology, Zhejiang University, Hangzhou, China}
\icmlaffiliation{ntu}{School of Computer Science and Engineering, Nanyang Technological University, Singapore}
\icmlaffiliation{zjuc}{College of Control Science and Engineering, Zhejiang University, Hangzhou, China}

\icmlcorrespondingauthor{Bo An}{boan@ntu.edu.sg}
\icmlcorrespondingauthor{Gang Pan}{gpan@zju.edu.cn}

\icmlkeywords{Machine Learning, ICML}

\vskip 0.3in
]



\printAffiliationsAndNotice{}  

\begin{abstract}
Ad hoc teamwork requires an agent to cooperate with unknown teammates without prior coordination. Many works propose to abstract teammate instances into high-level representation of types and then pre-train the best response for each type. However, most of them do not consider the distribution of teammate instances within a type. This could expose the agent to  the hidden risk of \emph{type confounding}.  In the worst case, the best response for an abstract teammate type could be the worst response for all specific instances of that type. This work addresses the issue from the lens of causal inference. We first theoretically demonstrate that this phenomenon is due to the spurious correlation brought by uncontrolled teammate distribution.  Then, we propose our solution, CTCAT, which disentangles such correlation through an instance-wise teammate feedback rectification. This operation reweights the interaction of teammate instances within a shared type to reduce the influence of type confounding. The effect of CTCAT is evaluated in multiple domains, including classic ad hoc teamwork tasks and real-world scenarios. Results show that CTCAT is robust to the influence of type confounding, a practical issue that directly hazards the robustness of our trained agents but was unnoticed in previous works.  
\end{abstract}

\section{Introduction}
\label{sec:introduction}

Developing agents that can cooperate with teammates without prior coordination is a well-established problem in the community of multi-agent systems, usually known as ad hoc teamwork \cite{bowling05aaai,stone10aaai}. Consider a scenario where an agent is required to cooperate with some unknown teammates in an emergency rescue operation. Due to the urgent situation, the agent has no opportunity to negotiate with others about their division of work before the task begins. However, in order to achieve the common goal, the agent needs to cooperate with others effectively on the fly, without knowing their abilities. This forms a typical scenario of ad hoc teamwork. As agents proliferate in the real world and their functions become more diversified, the application of ad hoc teamwork is prevalent in many domains, such as online games \cite{canaan19hanabi},  human-computer interactions \cite{suriadinata21hci} and visual navigation \cite{wang21visualnavigation}.

For practical ad hoc teamwork tasks where most teammates only appear occasionally, learning the best response for each individual becomes too costly \cite{rovatsos2002towards}. Instead, many works address the problem through the framework of type-based approach \cite{mirsky22survey}. These works strive to abstract teammate behaviors into high-level representation of types. During deployment, the agent represents the unknown teammate with a most likely type and then employs the strategy pre-trained for that type. This approach works well when the type suitably represents the teammate, and the agent's response for that type is properly trained. However, we notice that most existing works do not consider the distribution of teammate instances when pre-training the agent's best responses. Instead, they are obtained upon the aggregated behaviors of instances within a certain type. This could expose the agent to the hidden risk of \emph{type confounding}, a problem that was unnoticed in previous works but would directly hazard the robustness of our agent. In the worst case, the best response for an abstract teammate type becomes the worst response for all specific instances belonging to that type. 

\begin{figure}[t]
    \vskip 0.2in
    \begin{center}
        \centerline{\includegraphics[width=\columnwidth]{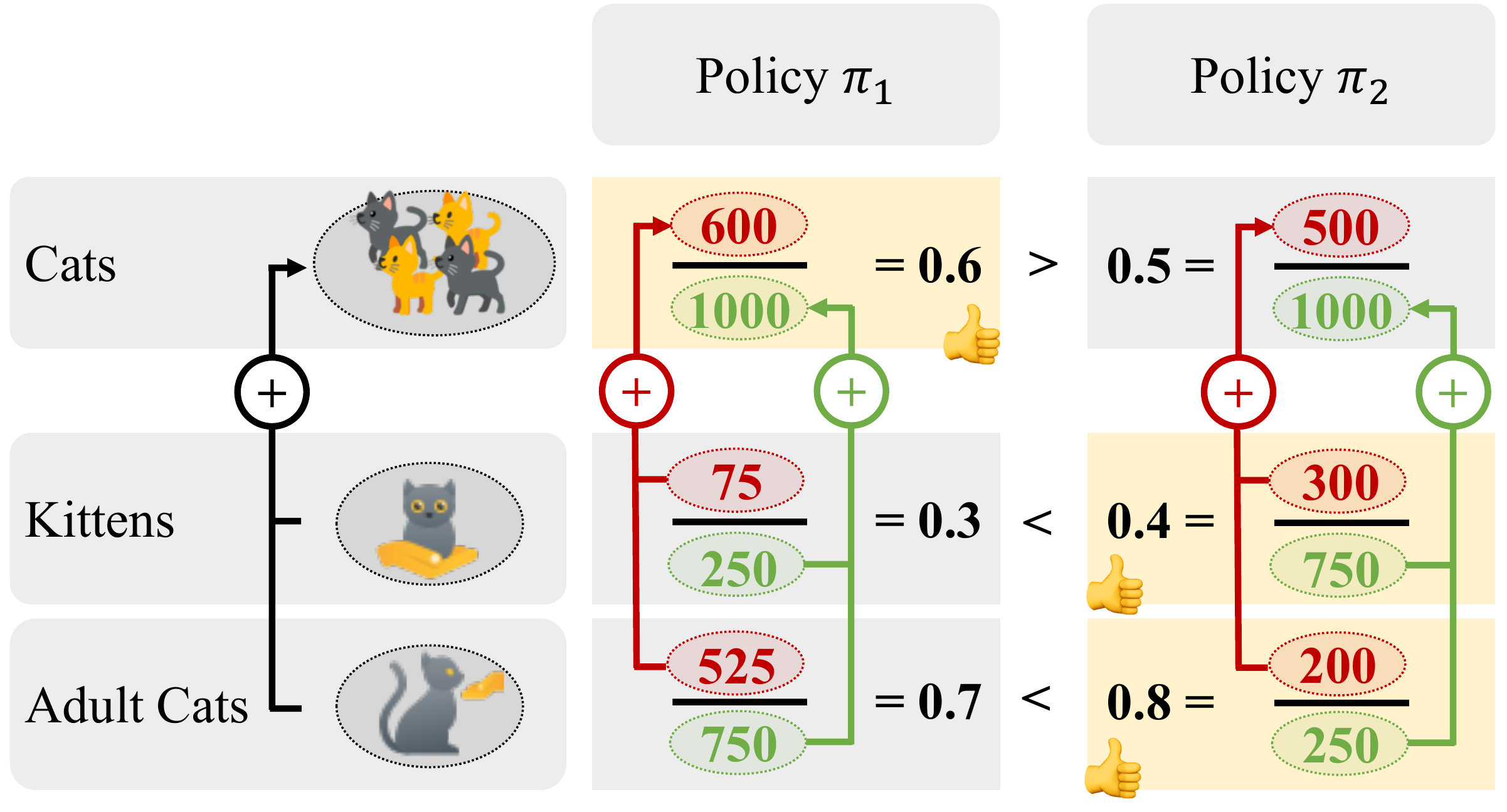}}
        \caption{
            An example of \emph{type confounding} in ad hoc teamwork. For the general cat-type pets, $\pi_1$ is the best response. However, for all instances of this type, $\pi_2$ is more favorable than $\pi_1$. 
        }
        \label{fig:simpson}
    \end{center}
    \vskip -0.2in
\end{figure}

To demonstrate the risk of type confounding more intuitively, we provide an example in Figure \ref{fig:simpson}. Suppose Alice runs a pet boarding shop where her business is to temporarily take care of pets from busy owners. Most pets are new to Alice, so it forms a kind of ad hoc teamwork between her and all the pets. Alice assumes that in unfamiliar environments, some pets need more company while others prefer to be alone. Therefore, she prepares two policies ($\pi_1$ and $\pi_2$) and starts to infer the best response for each type of pet. Alice observes that for pets of type \textit{cat} in her shop, 600 out of 1,000 customers give positive feedback when she applies $\pi_1$, while only 500 out of 1,000 customers give positive feedback to $\pi_2$. Thus, Alice believes $\pi_1$ is the best response for future customers with cats. One day, Alice wants to examine the impact of cat age on her policies, so she divides all cats into two groups: kittens and adult cats.  Surprisingly, for both groups, $\pi_2$ is more favorable than $\pi_1$. This  suggests the initial choice of Alice was problematic: the best response for cat-type pets is the worst response for all instances forming that type.\footnote{This phenomenon is known as Simpson's paradox.} This example demonstrates the risk of type confounding if the agent does not consider the distribution of teammate instances.\footnote{The term \textit{teammate instance} refers to teammate individuals belonging to a certain type, featured by specific properties (cat age in this example). It is named as such because it represents the instantiation of teammates belonging to an abstract type. } 

This work analyzes type confounding from the lens of causal inference \cite{pearl09causality}. We first theoretically demonstrate that this phenomenon is due to spurious correlation brought by uncontrolled teammate distribution,  which forms a confounder between the agent's policy and the cooperation outcome.  Briefly speaking, a confounder is a factor that simultaneously influences the cause and effect. In the context of ad hoc teamwork, the agent's policy (the cause) is adjusted during its interaction with the teammate, and they together determine the cooperation outcome (the effect). Therefore, the teammate distribution forms a confounder when the agent pre-trains the best response towards a certain type. If the confounder is not handled properly, it will easily introduce spurious correlations between cause and effect, making the learned best response unreliable. We name this phenomenon \emph{type confounding} due to its close relationship with the teammate type. This problem cannot be directly solved by regular practices 
such as collecting more data, since it is caused by the intrinsic structure of the dependency graph which is invariant to data size.

Based on this finding, we propose our solution CTCAT (Controlling Type Confounding in Ad hoc Teamwork) to address the issue of type confounding. Specifically, CTCAT disentangles the spurious correlation between the agent's policy and the cooperation outcome through an instance-wise teammate feedback rectification, which is derived by performing causal inference over the distribution of optimal cooperation outcomes. This operation reweights the interaction of teammate instances within a shared type to make them align with an ideally unbiased distribution. With this procedure, the spurious correlation between the agent's policy and the cooperation outcome is untangled,  which reduces the influence of type confounding. We evaluate the effect of CTCAT in multiple domains, including classic ad hoc teamwork tasks and real-world scenarios. Results show that our solution is robust to type confounding, a practical issue that directly hazards the robustness of our trained agents but was unnoticed in previous works. To our knowledge, CTCAT is the first work to (1) unveil the existence of type confounding in ad hoc teamwork, and (2) propose a causality-based solution to reduce its influence.

\section{Related Work}
\label{sec:related_works}

This section discusses related works of ad hoc teamwork. We divide relevant research into solutions with handcrafted teammate types and learning-based ones.

\subsection{Handcrafted Teammate Types} 

Many works on ad hoc teamwork directly train the agent's best response towards a set of handcrafted teammate types. For example, 
Albrecht and Ramamoorthy (\citeyear{albrecht13hba}) proposed a solution called HBA, which modeled ad hoc teamwork within the framework of Bayesian game \cite{harsanyi1967games}. They applied the solution to tasks with real humans, and the agent's policy was trained on teammate types designed by experts. The PLASTIC model, proposed by Barrett et al. (\citeyear{barrett17plastic}), suggested inferring the teammate types through a dynamic procedure of Bayesian posterior approximation. They enabled the model to work in a synthetic soccer game, where different policies were trained to cooperate with dedicated soccer teams \cite{hausknecht2016half}. Chen et al. (\citeyear{chen20aateam}) designed an attention network to perform teammate type inference, which incorporated the temporal information flexibly. Nevertheless, their solution was still based on pre-training best responses over a static set of teammate types, which they chose to inherit the setting of Barrett et al. (\citeyear{barrett17plastic}). Ravula et al. (\citeyear{ravula19convcpd}) proposed a changing point detector to monitor the time point at which the teammate changes its behavior patterns. The teammate types in their work were obtained by a heuristic search algorithm to find the optimal path \cite{hart1968formal}. A common characteristic of these works is that their teammate types are abstractions of manually specified behaviors, which corresponds to a distribution of teammate instances. However, when the agent pre-trains the best response for a certain type, most works do not consider the distribution of instances within that type. This opens the door of type confounding, which hazards the robustness of learned strategies.

\begin{figure}[t]
    \vskip 0.2in
    \begin{center}
        \centerline{
            \includegraphics[width=\columnwidth]{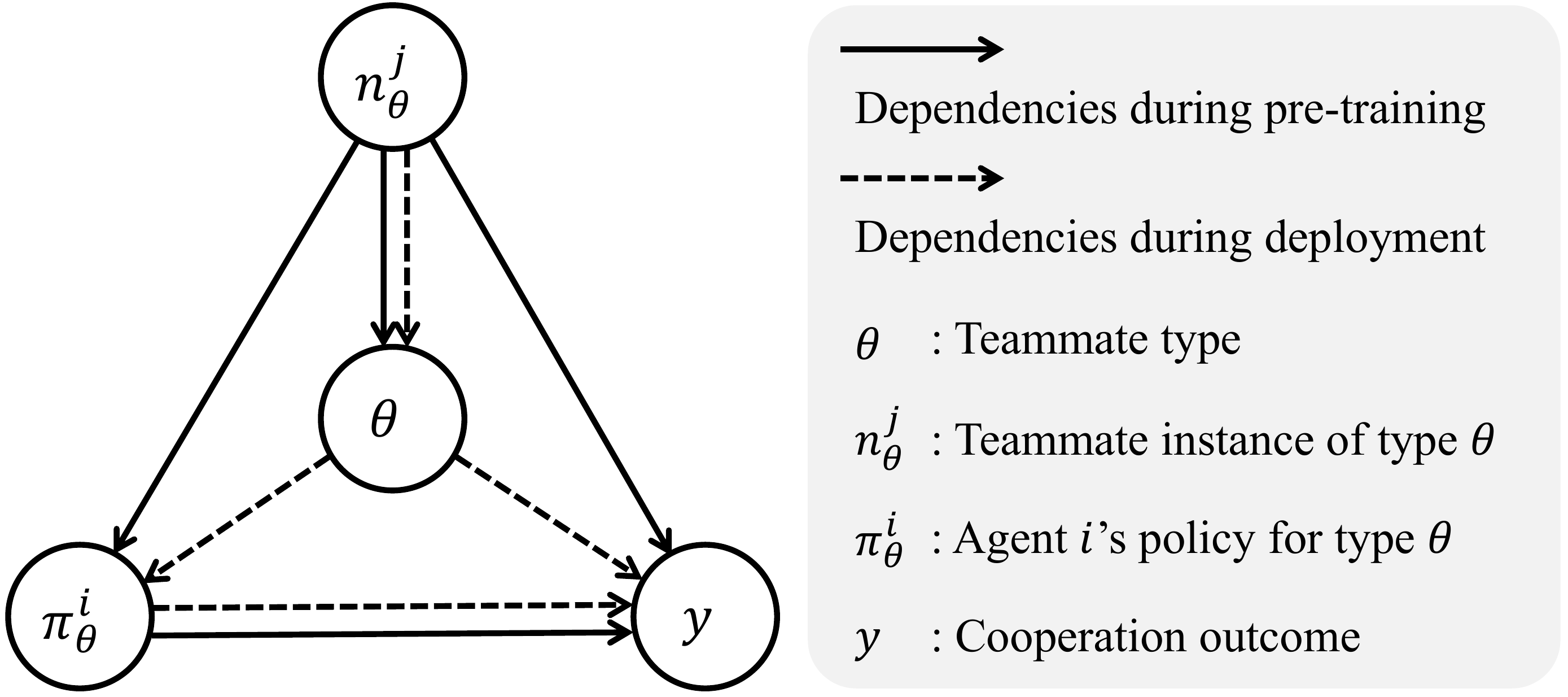}
        }
        \caption{The dependency graph of ad hoc teamwork during pre-training (solid arrows) and deployment (dash arrows). }
        \label{fig:dag}
    \end{center}
    \vskip -0.2in
\end{figure}

\subsection{Learning-Based Teammate Types}
To avoid the labor of manually designing teammate types, many works propose to model ad hoc teamwork in a more flexible approach. For example, Papoudakis et al. (\citeyear{papoudakis21liam}) proposed to reconstruct the teammate's behaviors by the agent's local observation through an encoder-decoder network,  where similar behaviors were encoded into adjacent embedding vectors. The agent's response was then generated with a policy conditioned on the inferred teammate embedding. Zintgraf et al. (\citeyear{zintgraf21meliba}) adopted a similar approach, where their teammate embeddings were obtained by a sequential variational auto-encoder. Gu et al. (\citeyear{gu2021online}) proposed to introduce an information-based regularizer to derive proxy representations of the teammate. This proxy representation was then used to generate hyper networks which guide the agent's best response. Melo \& Sardinha (\citeyear{melo16}) proposed to simultaneously identify the teammate's strategy and the task to be completed, which leads to policies that are more task-oriented. Some works expand the teammate types into larger space. For instance, Xing et al. (\citeyear{xing2021learning}) proposed to generate the teammate with an entropy regularizer. Rahman et al. (\citeyear{rahman21openaht}) proposed to solve ad hoc teamwork in an open environment, and the teammate type was captured with a graph neural network. Although these works adopt more advanced approaches to represent the teammate types, their best responses towards a given teammate type are directly trained upon the set of behaviors associated with that type. In this procedure, the distribution of teammate instances is usually neglected. Therefore, the problem of type confounding still exists, since it is rooted in the structure of ad hoc teamwork's dependency graph which is invariant to either the format of type representation or the capacity of type space. In consequence, a new solution is required to address the issue of type confounding, and our work is the first attempt to fill this blank.  

\section{Preliminaries}
\label{sec:preliminaries}

This section provides preliminaries of our work. We first review the framework of stochastic Bayesian game \cite{harsanyi1967games,albrecht16stochasticBayesian}, which models the interaction  between our controlled agent with an unknown teammate. Then, we introduce the dependency graph of ad hoc teamwork during pre-training and deployment (depicted in Figure \ref{fig:dag}), which contains the cause of type confounding. 

\subsection{Stochastic Bayesian Game}\label{subsec:bayesian_game}

The framework of stochastic Bayesian game is suitable to model ad hoc teamwork since it is dedicated to scenarios where agents interact with some unknown players. Specifically, a stochastic Bayesian game can be defined as a tuple of  $\left<  \mathcal{N}, \mathcal{S}, \left\{ \mathcal{A}^n \right\},  \Theta, P, R, \gamma \right> $, where 
$\mathcal{N} $ is the set of all agents $n \in \mathcal{N}$, and $\mathcal{N}_{\theta}$ denotes the set of agents with type $\theta$. 
$\mathcal{S}$ is the set of all valid states $\mathbf{s} \in \mathcal{S}$. 
$\mathcal{A}^n$ is the set of valid actions ${a}^n \in \mathcal{A}^n$ for agent $n$ and $\mathcal{A} = \times_{n \in \mathcal{N}} \mathcal{A}^n$ is the set of joint action $\mathbf{a}$. 
$\Theta$ is the set of all possible types for the teammate, where each type is represented as $\theta \in \Theta$. 
$P: \mathcal{S} \times \mathcal{A} \times \mathcal{S} \rightarrow [0, 1]$ is the transition function.  
$R: \mathcal{S} \times \mathcal{A} \rightarrow \mathbb{R}$ is the reward function, which is shared by all agents since we focus on cooperative tasks.  
$\gamma \in (0, 1]$ is the discount factor for future rewards. 

Without loss of generality, we use the superscript $i$ to denote our controlled agent, which needs to cooperate with the unknown teammate $j$ without prior coordination. The teammate's type representation $\theta \in \Theta$ is abstracted from similar teammate behaviors, and all instances with a common teammate type share an identical best response $\pi_\theta^i$. 
This enables the learned policy to generalize to teammates unseen but with similar type representations. 
The goal of our agent is to design conditional policy $\pi^i_\theta$ which maximizes the cumulative reward $y({\theta} )= \mathbb{E}[ \sum_{t=0}^{T} \gamma^t r_t  \mid \theta ]$ given the inferred teammate type $\theta$, where $t$ is the time step and $T$ is the time horizon. We use $y^\star$ to denote the optimal value of $y$. 

\subsection{The Dependency Graph}\label{subsec:dependency_graph}

The following analysis is based on the dependency graph of ad hoc teamwork presented in Figure \ref{fig:dag}. In this figure, directed arrows represent dependency relationships among the teammate's type $\theta$, its instance $n_\theta^j$, the agent's policy $\pi^i_\theta$ and the cooperation outcome $y$. During pre-training, the agent's policy $\pi^i_\theta$ is adjusted with respect to 
its interaction with the teammate $n_\theta^j$, and they together determine the cooperation outcome $y$. Meanwhile, the agent abstracts teammate instances into a high-level type representation $\theta$, which aggregates similar teammate behaviors with a shared best response. This procedure can be performed manually \cite{barrett17plastic,ravula19convcpd} or learned automatically in a data-driven fashion \cite{papoudakis21liam,gu2021online}. Although the implementation varies in different works, they are all based on a common assumption: teammates with the same type representation $\theta$ share a common best response $\pi^i_\theta$. This assumption is fundamental in ad hoc teamwork, which ensures that the learned policy can be generalized to future unknown teammates with similar type representations \cite{bowling05aaai}. During deployment, the agent first represents the unknown teammate with a most likely type and then employs the best response conditioned on that type. The inferred teammate type and the agent's policy based on that type determine the quality of cooperation outcome. This procedure exempts the agent from training independent responses for all teammates, which is too costly to be implemented \cite{rovatsos2002towards,xing22tinylight}. The causal relationships presented in this graph are in accordance with many previous works on ad hoc teamwork \cite{mirsky22survey}.

From the dependency graph, we can observe that the distribution of teammate $n_\theta^j$ is a common cause for both the agent's policy $\pi_\theta^i$ and the cooperation outcome $y$ when the agent pre-trains its best responses. This indicates that the teammate distribution forms a confounder between them.\footnote{This work assumes that teammate instance ($n_\theta^j$) is the only confounder and it is fully accessible. This assumption is often referred as \textit{unconfoundedness} in the literature of causal inference \cite{rubin1978bayesian}.  }  If the confounder is not handled properly, it will induce bias into the learned best response, distorting the causal relationship between $\pi_\theta^i$ and $y$. In the worst case, the unfavorable action becomes spuriously more favorable, which makes the agent's policy $\pi^i_\theta$ no longer reliable, even though the inferred teammate type $\theta$ is correct. This problem cannot be directly solved with richer data, since it is caused by the intrinsic structure of the dependency graph which is invariant to the data size.\footnote{This problem is different from covariate shift. Covariate shift is caused by different distributions between training and test sets, leading to sample selection bias. Meanwhile, type confounding is caused by an uncontrolled confounder which affects both the cause and effect, leading to confounding bias. }

\section{CTCAT}
\label{sec:method}

This section provides our proposed solution. We first present the ideal distribution of optimal outcome $y^\star$ in Section \ref{subsec:ideal_dist}, which is unbiased but is unfortunately inaccessible. Meanwhile, we present the practical distribution of $y^\star$ in Section \ref{subsec:practical_dist}, which is accessible but could be possibly biased due to type confounding. Our proposed instance-wise teammate feedback rectification is covered in Section \ref{subsec:control_for_confounding}, which controls the influence of  type confounding by approximating a distribution that is both unbiased and accessible.

\subsection{The Ideal Distribution of $y^\star$}
\label{subsec:ideal_dist}

We use the $do$-operator to proceed with our analysis, which is a standard tool provided by the literature of causal inference \cite{pearl09causality}. Briefly speaking, the $do$-operator corresponds to the intervention of our interested variable in order to verify its causal correlation on the outcome variable. Specifically, we use $p\left( y^\star \mid do\left( \pi^i_\theta \right), \theta  \right) $ to denote the probability of reaching the optimal outcome $y^\star$ when the agent is interfered to adopt policy $\pi^i_\theta$ for a given inferred teammate type $\theta$. For a pair of policies $\pi_{\theta,1}^i$ and $\pi_{\theta,2}^i$, the $do$-operator allows us to conclude with causality guarantee that for the given inferred teammate type $\theta$, $\pi_{\theta,1}^i$ is better than $\pi_{\theta,2}^i$ if $p (y^\star \mid do( \pi_{\theta,1}^i ), \theta ) > p( y^\star \mid do( \pi_{\theta,2}^i ), \theta )$. With the law of total probability, $p( y^\star \mid do( \pi_\theta^i ), \theta  ) $ can be transformed into the following form: 
\begin{equation}
    \label{eq:ideal_total_prob}
    \begin{aligned} 
        & p\left( y^\star  \mid do\left(  \pi_{\theta}^i \right), \theta \right) 
        = \\
        & ~~~~\sum_{n_\theta^j}
        p( y^\star \mid  \underbrace{do\left( \pi_\theta^i \right)}_{\text{1st }do\text{-op}}, \theta, n_\theta^j  ) \cdot 
        p(  n_\theta^j \mid \underbrace{do \left( \pi_\theta^i \right)}_{\text{2nd }do\text{-op}}, \theta )
    \end{aligned} 
\end{equation}
The $do$-operator is a conceptual operation and should be replaced with observable estimands to identify its real value. In the following, we perform identification on the two $do$-operators in the right-hand side of Eq. (\ref{eq:ideal_total_prob}) sequentially. 

\paragraph{Identification of the first $do$-operator.}

The first $do$-operator requires us to identify the causal correlation from $\pi_\theta^i$ to $y^\star$. The correlations from $\pi_\theta^i$ to  $y^\star$ are conveyed by paths that either flow out of $\pi_\theta^i$ (front-door paths) or flow into $\pi_\theta^i$ (back-door paths). In our dependency graph, the unique back-door path from $\pi_\theta^i$ to  $y^\star$ is $\pi_\theta^i \leftarrow n_\theta^j \rightarrow y^\star$, which is blocked when the variable $n_\theta^j$ is given. Therefore, the remaining correlation between $\pi^i_\theta$ and $y^\star$ is the front-door path $\pi_\theta^i \rightarrow y^\star$, which is exactly the causal correlation between them. Thus, the $do$-operator can be directly removed from the first term of Eq. (\ref{eq:ideal_total_prob}), which gives us:
\begin{equation}
    \label{eq:ideal_1}
    p\left(  
    y^\star \mid do\left(  \pi_\theta^i \right), \theta, n_\theta^j
    \right) 
    = 
    p\left(
    y^\star \mid \pi_\theta^i, \theta, n_\theta^j
    \right)
\end{equation}
We can further remove $\theta$ from the conditional variables in Eq. (\ref{eq:ideal_1}). This is made possible by the fact that the 
correlation between $\theta$ and $y^\star$ is formed by paths of  $\theta \leftarrow n_\theta^j \rightarrow y^\star$ and $\theta \leftarrow n_\theta^j \rightarrow \pi^i_\theta \rightarrow y^\star$, which are both blocked if the variable $n_\theta^j$ is given. Therefore, $\theta$ and $y^\star$ are conditionally independent when $n_\theta^j$ is provided, which gives us:  
\begin{equation}
    \label{eq:ideal_3}
    p\left(
    y^\star \mid \pi_\theta^i, \theta, n_\theta^j
    \right)
    = 
    p\left(
    y^\star \mid \pi_\theta^i, n_\theta^j 
    \right)
\end{equation}

\paragraph{Identification of the second $do$-operator.} 
The second $do$-operator requires us to identify the causal correlation from $\pi_\theta^i$ to $n_\theta^j$. If a variable is intervened by the $do$-operator, all its upstream dependencies are removed from the graph since this variable is fully controlled by the $do$ operation. Under this condition, the unique path from $\pi_\theta^i$ to $\theta$ is $\pi_\theta^i \rightarrow y^\star \leftarrow \theta$, where $y^\star$ forms a collider between $\pi_\theta^i$ and $\theta$ and blocks the dependency flow within this path. Therefore, the second $do$-operator can be removed directly, which gives us: 
\begin{equation} 
    \label{eq:ideal_2}
    p\left( n^j_\theta \mid do \left( \pi^i_\theta \right), \theta  \right)  
    = 
    p\left(  n_\theta^j \mid \theta \right)
\end{equation}
We summarize the conclusions of Eq. (\ref{eq:ideal_total_prob}) - (\ref{eq:ideal_2}) formally into the following theorem: 
\begin{theorem}
    \label{prop:ideal}
    With the dependency graph of ad hoc teamwork, the value of $p\left( y^\star \mid do\left( \pi_\theta^i \right), \theta  \right) $ can be identified as: 
    \begin{equation}
        \label{eq:prop_1}
        p\left(  y^\star \mid do\left( \pi_\theta^i \right), \theta  \right)  
        = 
        \sum _ {n_\theta^j}
        p \left(
        y^\star \mid \pi^i_\theta, n_\theta^j 
        \right)
        \cdot 
        p \left(
        n_\theta^j \mid \theta 
        \right)
    \end{equation}
\end{theorem}
    
The intuition behind Theorem \ref{prop:ideal} is that the unbiased value of $y^\star$ should be obtained by integrating the value of $p( y^\star \mid \pi_\theta^i, n_\theta^j)$, weighted by the probability of having $n_\theta^j$ when the teammate type is fixed to $\theta$.  The first term $p(y^\star \mid \pi_\theta^i, n_\theta^j)$ represents the likelihood of being optimal when the agent employs policy $\pi_\theta^i$ to cooperate with teammate $n_\theta^j$. The second term $p(n_\theta^j \mid \theta)$ denotes the distribution of teammate instances with the given type $\theta$, which is invariant to the choice of $\pi_\theta^i$. Unfortunately, $n_\theta^j$ and $\pi_\theta^i$ are not completely independent in the dependency graph. This makes the value of $p(n_\theta^j \mid \theta)$ (and therefore $p(y^\star \mid do(\pi_\theta^i), \theta)$) not directly accessible, even though it forms an unbiased estimation of the optimal outcome from the lens of causal inference. 
    
\subsection{The Practical Distribution of $y^\star$}
\label{subsec:practical_dist}

In practice, the agent's best response towards a given teammate type is usually constructed by training the optimal policy toward aggregated behaviors of teammate instances belonging to this type. This procedure corresponds to the following distribution of $y^\star$: 
\begin{equation}
    \label{eq:dist_practical}
    \begin{aligned}
        p \left( 
        y^\star \mid \pi^i_\theta, \theta 
        \right)
        = & 
        \sum_{n_\theta^j}
        p \left(
        y^\star \mid \pi_\theta^i, \theta , n_\theta^j  
        \right)
        \cdot 
        p \left(
        n_\theta^j \mid \pi^i_\theta, \theta
        \right) 
        \\ 
        = & 
        \sum_{n_\theta^j}
        p \left(
        y^\star \mid \pi_\theta^i, n_\theta^j  
        \right)
        \cdot 
        p \left(
        n_\theta^j \mid \pi^i_\theta, \theta
        \right) 
    \end{aligned}
\end{equation}
where the first line is derived by the law of total probability, and the second line is derived by the fact that $\theta$ and $y^\star$ are conditionally independent when the variable of $n_\theta^j$ is given (mentioned in Eq. (\ref{eq:ideal_3})). This gives us the following theorem: 
\begin{theorem}
    \label{prop:practical}
    With the dependency graph of ad hoc teamwork, the value of $p\left( y^\star \mid  \pi^i_\theta, \theta  \right)$ can be represented as: 
    \begin{equation}\label{eq:prop2}
        p \left(
        y^\star \mid \pi_\theta^i, \theta 
        \right)
        = 
        \sum_{n_\theta^j}
        p\left(
        y^\star \mid \pi_\theta^i, n_\theta^j 
        \right) 
        \cdot 
        p \left(
        n_\theta^j \mid \pi_\theta^i, \theta
        \right)
    \end{equation}    
\end{theorem}
The value of Eq. (\ref{eq:prop2}) is accessible, since it can be directly obtained by learning from past experiences. The intuition behind Theorem \ref{prop:practical} is similar to Theorem \ref{prop:ideal}. The only difference is that the value of $p(y^\star \mid \pi^i_\theta, n_\theta^j)$ is now weighted by $p(n_\theta^j \mid \pi_\theta^i, \theta)$, the distribution of teammate instances when the agent adopts policy $\pi_\theta^i$  with the given type $\theta$. This is a more practical estimand of the teammate distribution, since in reality there are many factors that can lead to teammate instances being unevenly distributed under different agent policies, such as curiosity-based exploration \cite{ecoffet2021first}, sampling complexity \cite{yang21aaai2} or offline replay buffers \cite{agarwal20offlineRL,gu22offline}. It makes the agent's estimated distribution $y^\star$ biased, which distorts the causal relationship between $\pi_\theta^i$ and $y^\star$. In the worst case, the best response for a teammate type $\theta$ becomes the worst response for all instances of this type. 

\subsection{Instance-wise Teammate Feedback Rectification}
\label{subsec:control_for_confounding}

Comparing Theorem \ref{prop:ideal} with \ref{prop:practical}, we can conclude that the deviation between the ideal and practical distribution of $y^\star$ is caused by the discrepancy between weighting terms $p(n_\theta^j \mid \theta)$ and  $p(n_\theta^j \mid \pi_\theta^i, \theta)$. The former term denotes the marginalized distribution of teammate instances which is shared by all agent policies. The latter term denotes the conditional distribution of teammate instances when the agent adopts a specific policy $\pi_\theta^i$. With this observation, we can rectify the bias by re-weighting the importance of each teammate feedback in an instance-wise manner. Specifically, we insert into Eq. (\ref{eq:dist_practical}) a weighting factor $w$: 
\begin{equation}
    \label{eq:with_weight}
    \begin{aligned}
        & \hat{p} \left(
        y^\star \mid \pi_\theta^i, \theta 
        \right)
        = 
        \\ 
        & ~~~~
        \sum_{n_\theta^j}
        p\left(
        y^\star \mid \pi_\theta^i, n_\theta^j 
        \right) 
        \cdot 
        p \left(
        n_\theta^j \mid \pi_\theta^i, \theta
        \right) 
        \cdot 
        w\left(
        \pi_\theta^i, n_\theta^j, \theta  
        \right)
    \end{aligned}
\end{equation}    
where: 
\begin{equation}\label{eq:weight}
    w\left( \pi_\theta^i, n_\theta^j, \theta  \right) = {
        p \left(  n_\theta^j \mid \theta \right)
    } / { 
        p \left(  n_\theta^j \mid \pi_\theta^i, \theta \right)
    }
\end{equation}
        
The original form of Eq. (\ref{eq:weight}) is still hard to estimate due to the existence of inaccessible term $p (  n_\theta^j \mid \theta )$. Fortunately, this term can be eliminated with some transformations on the denominator of Eq. (\ref{eq:weight}) based on the Bayesian rule: 
\begin{equation}\label{eq:bayesian}
    p\left( n_\theta^j \mid \pi_\theta^i, \theta \right) 
    = 
    \frac{
        p \left( n_\theta^j \mid \theta  \right) \cdot  p \left( \pi_\theta^i \mid \theta, n_\theta^j  \right)
    }{
        p \left( \pi_\theta^i \mid \theta  \right)
    }
\end{equation}

\begin{table*}[t]
    \caption{The distribution of selected policies on different teammate distributions. For all teammate instances, $\pi_1$ is the best response. However, $\pi_2$, $\pi_3$ and $\pi_4$ become spuriously more favored respectively if the teammate distributions are distorted in different ways. } 
    \label{tab:overall_result} 
    \begin{center}
        \begin{tabular}{c c cccc c cccc}
            \toprule
            \begin{minipage}{0.25\linewidth} 
                    \includegraphics[width=\linewidth,height=0.015\textheight]{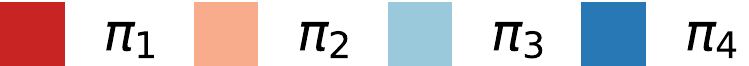} 
            \end{minipage}
            
            && \multicolumn{4}{c}{\begin{minipage}{0.3\linewidth} 
                    \centering{(1) The data are evenly distributed. }
            \end{minipage}}
            && \multicolumn{4}{c}{\begin{minipage}{0.3\linewidth} 
                    \centering{(2) $\pi_2$ is spuriously more favored. }
            \end{minipage}} \\ 
            \midrule 
            FIAM \cite{papoudakis21liam}
            && \multicolumn{4}{c}{\begin{minipage}{0.3\linewidth} 
                        \includegraphics[width=\linewidth,height=0.015\textheight]{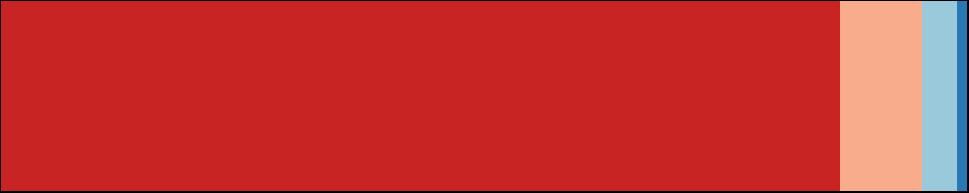} 
            \end{minipage}} 
            && \multicolumn{4}{c}{\begin{minipage}{0.3\linewidth} 
                    \includegraphics[width=\linewidth,height=0.015\textheight]{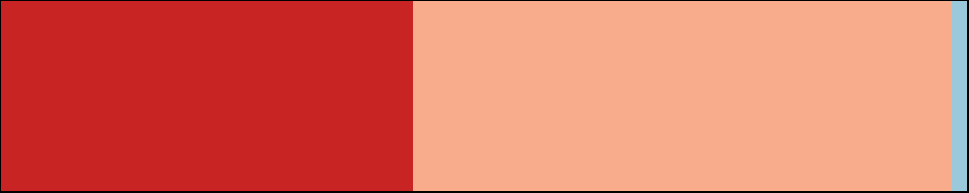} 
            \end{minipage}}  \\
            
            LIAM \cite{papoudakis21liam}      
            && \multicolumn{4}{c}{\begin{minipage}{0.3\linewidth} 
                    \includegraphics[width=\linewidth,height=0.015\textheight]{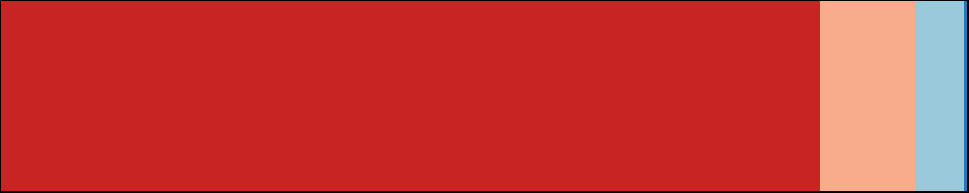} 
            \end{minipage}} 
            && \multicolumn{4}{c}{\begin{minipage}{0.3\linewidth} 
                    \includegraphics[width=\linewidth,height=0.015\textheight]{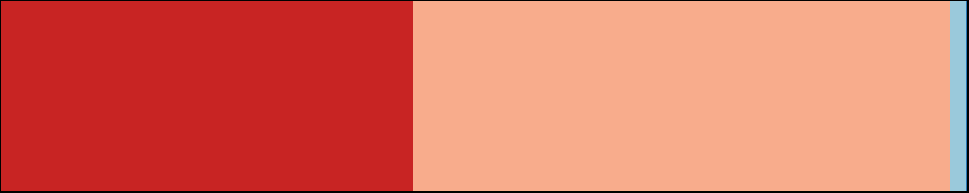} 
            \end{minipage}}  \\
            
            MELIBA \cite{zintgraf21meliba}
            && \multicolumn{4}{c}{\begin{minipage}{0.3\linewidth} 
                    \includegraphics[width=\linewidth,height=0.015\textheight]{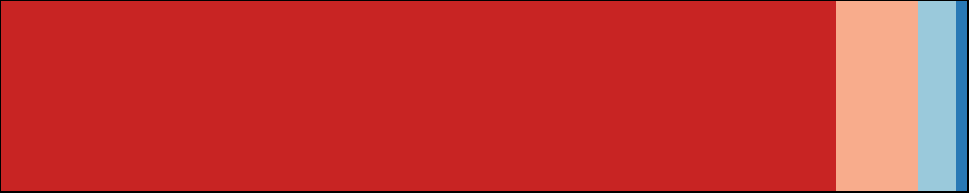} 
            \end{minipage}} 
            && \multicolumn{4}{c}{\begin{minipage}{0.3\linewidth} 
                    \includegraphics[width=\linewidth,height=0.015\textheight]{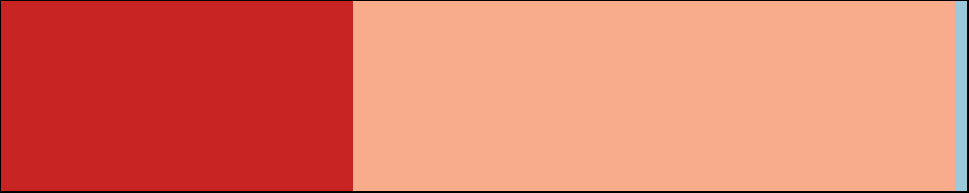} 
            \end{minipage}}  \\
            
            ODITS \cite{gu2021online}
            && \multicolumn{4}{c}{\begin{minipage}{0.3\linewidth} 
                    \includegraphics[width=\linewidth,height=0.015\textheight]{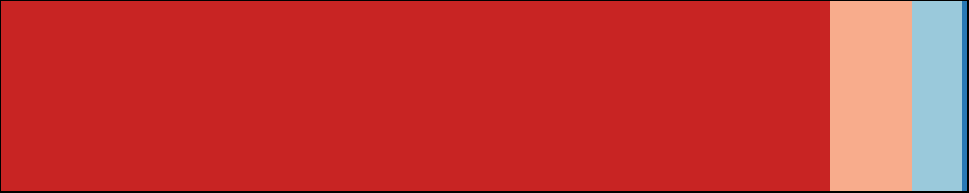} 
            \end{minipage}} 
            && \multicolumn{4}{c}{\begin{minipage}{0.3\linewidth} 
                    \includegraphics[width=\linewidth,height=0.015\textheight]{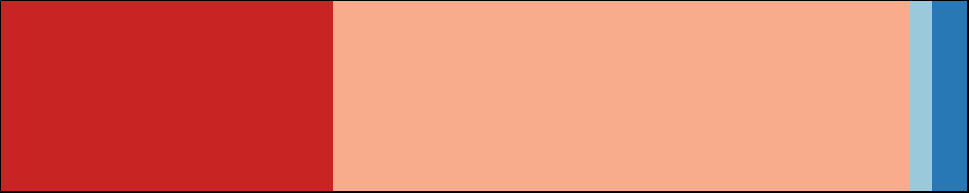} 
            \end{minipage}}  \\
            
            \midrule
            CTCAT (ablation)      
            && \multicolumn{4}{c}{\begin{minipage}{0.3\linewidth} 
                    \includegraphics[width=\linewidth,height=0.015\textheight]{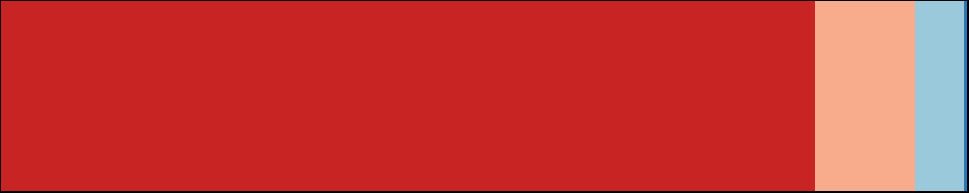} 
            \end{minipage}} 
            && \multicolumn{4}{c}{\begin{minipage}{0.3\linewidth} 
                    \includegraphics[width=\linewidth,height=0.015\textheight]{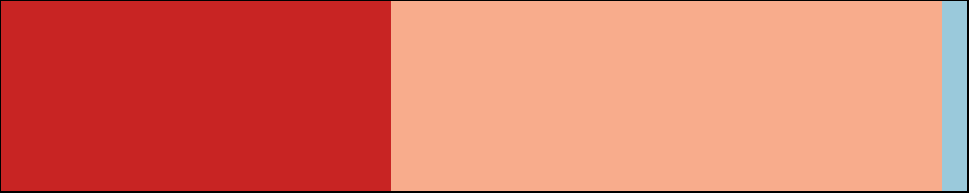} 
            \end{minipage}}  \\
            CTCAT   
            && \multicolumn{4}{c}{\begin{minipage}{0.3\linewidth} 
                    \includegraphics[width=\linewidth,height=0.015\textheight]{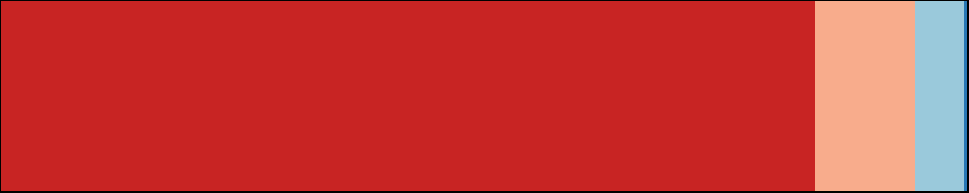} 
            \end{minipage}} 
            && \multicolumn{4}{c}{\begin{minipage}{0.3\linewidth} 
                    \includegraphics[width=\linewidth,height=0.015\textheight]{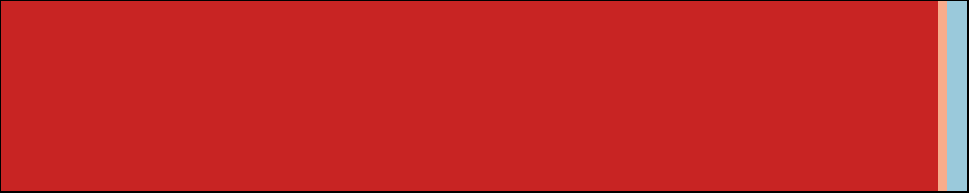} 
            \end{minipage}}  \\
            \bottomrule
            
            \toprule
            && \multicolumn{4}{c}{\begin{minipage}{0.3\linewidth} 
                    \centering{(3) $\pi_3$ is spuriously more favored.}
            \end{minipage}}
            && \multicolumn{4}{c}{\begin{minipage}{0.3\linewidth} 
                    \centering{(4) $\pi_4$ is spuriously more favored.}
            \end{minipage}} \\ 
            \midrule 
            FIAM  \cite{papoudakis21liam}
            && \multicolumn{4}{c}{\begin{minipage}{0.3\linewidth} 
                    \includegraphics[width=\linewidth,height=0.015\textheight]{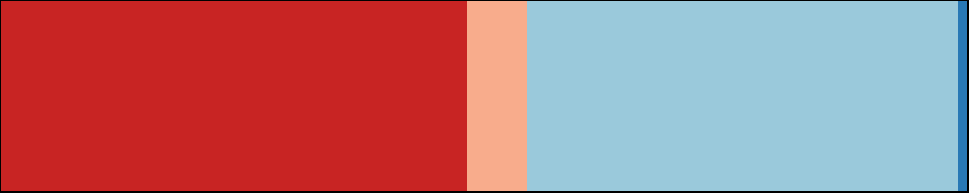} 
            \end{minipage}} 
            && \multicolumn{4}{c}{\begin{minipage}{0.3\linewidth} 
                    \includegraphics[width=\linewidth,height=0.015\textheight]{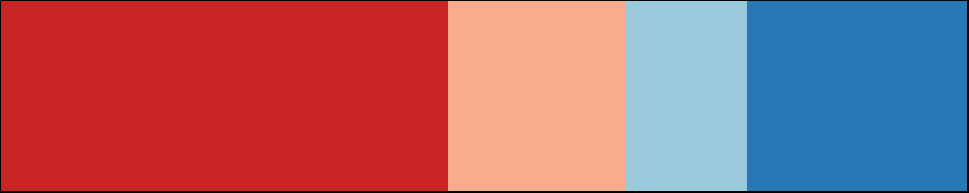} 
            \end{minipage}}  \\
            
            LIAM  \cite{papoudakis21liam}
            && \multicolumn{4}{c}{\begin{minipage}{0.3\linewidth} 
                    \includegraphics[width=\linewidth,height=0.015\textheight]{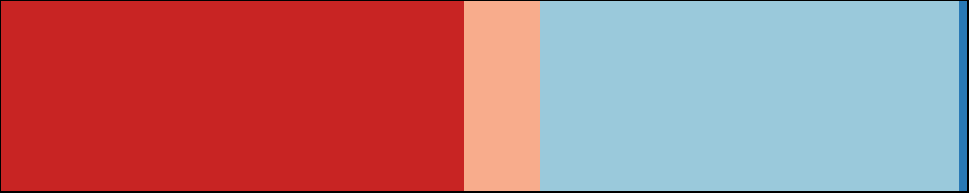} 
            \end{minipage}} 
            && \multicolumn{4}{c}{\begin{minipage}{0.3\linewidth} 
                    \includegraphics[width=\linewidth,height=0.015\textheight]{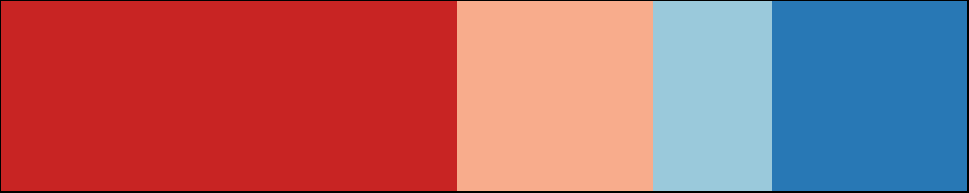} 
            \end{minipage}}  \\
            
            MELIBA \cite{zintgraf21meliba}
            && \multicolumn{4}{c}{\begin{minipage}{0.3\linewidth} 
                    \includegraphics[width=\linewidth,height=0.015\textheight]{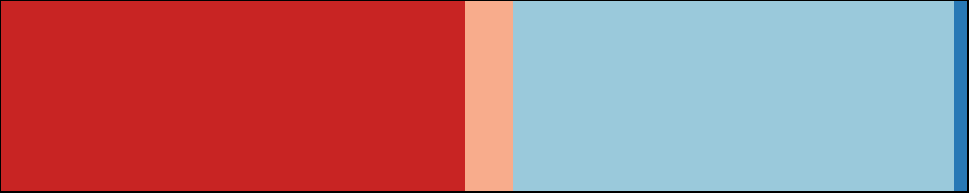} 
            \end{minipage}} 
            && \multicolumn{4}{c}{\begin{minipage}{0.3\linewidth} 
                    \includegraphics[width=\linewidth,height=0.015\textheight]{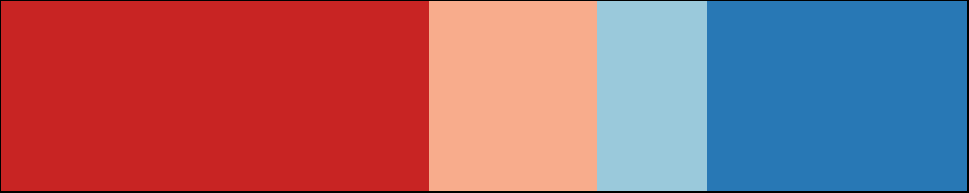} 
            \end{minipage}}  \\
            
            ODITS \cite{gu2021online}
            && \multicolumn{4}{c}{\begin{minipage}{0.3\linewidth} 
                    \includegraphics[width=\linewidth,height=0.015\textheight]{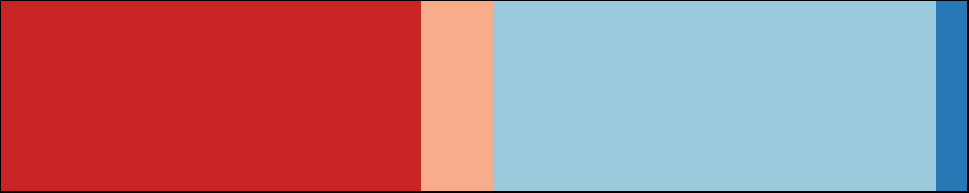} 
            \end{minipage}} 
            && \multicolumn{4}{c}{\begin{minipage}{0.3\linewidth} 
                    \includegraphics[width=\linewidth,height=0.015\textheight]{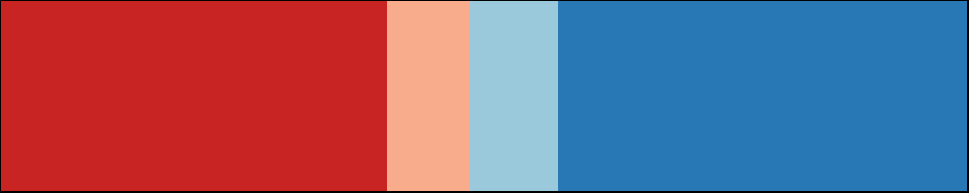} 
            \end{minipage}}  \\
            
            \midrule
            CTCAT (ablation)      
            && \multicolumn{4}{c}{\begin{minipage}{0.3\linewidth} 
                    \includegraphics[width=\linewidth,height=0.015\textheight]{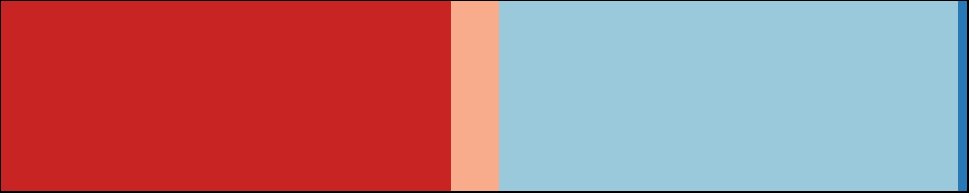} 
            \end{minipage}} 
            && \multicolumn{4}{c}{\begin{minipage}{0.3\linewidth} 
                    \includegraphics[width=\linewidth,height=0.015\textheight]{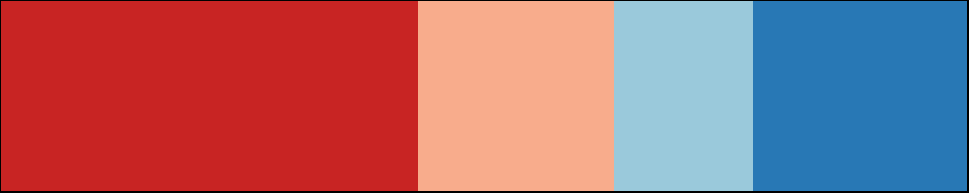} 
            \end{minipage}}  \\
            CTCAT   
            && \multicolumn{4}{c}{\begin{minipage}{0.3\linewidth} 
                    \includegraphics[width=\linewidth,height=0.015\textheight]{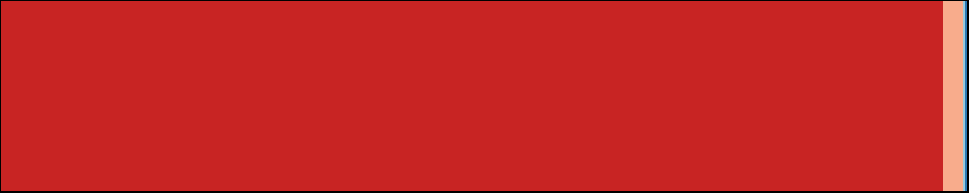} 
            \end{minipage}} 
            && \multicolumn{4}{c}{\begin{minipage}{0.3\linewidth} 
                    \includegraphics[width=\linewidth,height=0.015\textheight]{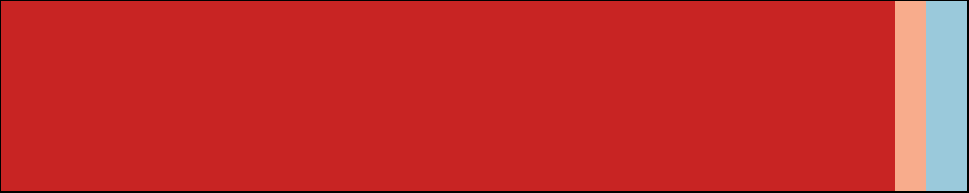} 
            \end{minipage}}  \\
            \bottomrule
        \end{tabular}
    \end{center}
\end{table*}

Combining Eq. (\ref{eq:with_weight}) - (\ref{eq:bayesian}), we have the following theorem:  
\begin{theorem}\label{prop:combine}
    With the dependency graph of ad hoc teamwork, the ideally unbiased distribution of $y^\star$ in Eq. (\ref{eq:prop_1}) can be represented as: 
    \begin{equation}
        \begin{aligned}
            \hat{p}\left(  
            y^\star \mid \pi_\theta^i, \theta
            \right)
            = & 
            \sum_{n_\theta^j} 
            p\left(
            y^\star \mid \pi_\theta^i, n_\theta^j 
            \right)  
            \cdot 
            p\left(
            n_\theta^j \mid \pi_\theta^i, \theta
            \right) \cdot \\
            &     
            \underbrace{
                p\left(
                \pi_\theta^i \mid \theta 
                \right)
                / 
                p\left(
                \pi_\theta^i \mid \theta, n_\theta^j 
                \right)
            }_{=w\left( \pi_\theta^i, n_\theta^j, \theta \right)}
        \end{aligned}
    \end{equation}
\end{theorem}
The weighting factor requires us to maintain two runtime variables when pre-training the best response $\pi_\theta^i$ for teammate type $\theta$. The first term $p(\pi^i_\theta \mid \theta )$ corresponds to a global policy distribution which is averaged over all teammate instances within the type $\theta$. This can be viewed as a prior aggregating the commonalities of all teammate instances. 
The second term is $p(\pi_\theta^i \mid \theta, n_\theta^j)$, which represents an individualized policy distribution that is dedicated to teammate instance $n_\theta^j$. This term measures the deviation induced by each individual during training. Due to the task requirement of ad hoc teamwork, the agent is expected to extract commonalities from similar teammate instances to form the best response toward a common teammate type. Against this background, the weighting factor rectifies the deviation of each teammate instance to make them align with an unbiased global policy distribution. This keeps the agent from being influenced by the teammate distributions,  which is the root cause of type confounding.

\section{Experiments}
\label{sec:experiments}

This section presents our experimental studies. We first quantitatively analyze the influence of type confounding in Section \ref{subsec:exp2} with a goal-based version of predator-prey,  where we design experiments to demonstrate the following two observations: (1) without rectification, the best response for a teammate type can be arbitrarily distorted by the distribution of teammate instances; (2) collecting more data cannot directly address the problem of type confounding, since it does not influence the dependencies of ad hoc teamwork. Then, we evaluate the effect of CTCAT in several real-world scenarios in Section \ref{subsec:exp1}.
            
\subsection{Goal-based Predator-prey} 
\label{subsec:exp2}

To quantitatively analyze the influence of type confounding, we propose an adapted version of the classic game predator-prey. 
This game is widely adopted in previous studies of ad hoc teamwork \cite{barrett17plastic,gu2021online}. 
The detailed implementation varies slightly in different works, but they all require multiple predators to simultaneously capture a prey within limited timesteps. In our implementation, there are four preys on the map and the range of movement for each prey is limited to an isolated subarea. Meanwhile, there are two predators and one of them is under our control. We first train a set of candidate agents with self-played goal-conditioned RL \cite{ghosh2021goalbased}, with the goals set as all feasible combinations of 4 preys, such as $\{1 \}$, $\{2, 4\}$, $\{1, 2, 4\}$.  The agent with a large set of goals (e.g. $\{1, 2, 4\}$) is likely to cover behaviors of agents that have a subset of its goals (e.g. $\{1\}$, $\{2, 4\}$). This setting allows us to configure agent behaviors by directly specifying its goals. 

To evaluate the impact of type confounding,  we compose a pseudo teammate type by specifying two instances ($K_1$ and $K_2$) as the known teammates to describe this type and one instance ($U$) as the unknown teammate for evaluation. Meanwhile, we prepare four candidate policies ($\pi_1$, $\pi_2$, $\pi_3$ and $\pi_4$) for the agent to choose from, among which  $K_1$, $K_2$ and $U$ share a common best response (here we let $\pi_1$ be the best response for all teammate instances). This is configured by letting $K_1$, $K_2$ and $U$ share a common target prey which is only specified in the goal list of $\pi_1$. Therefore, $K_1$, $K_2$ and $U$ have similar behaviors, and classifying them into a same type is reasonable. This setting is common in many works on ad hoc teamwork \cite{mirsky22survey}, where an agent is trained with a set of prepared teammates and then evaluated on some unknown instances. However, the agent's action is now restricted to choosing a suitable  policy from the set of candidates provided in advance. This setting enables us to maintain a common best response among all baselines, which makes quantitative analyzing the impact of type confounding feasible. 
            
We adopt four state-of-the-art baselines on ad hoc teamwork for comparison, including FIAM \cite{papoudakis21liam}, LIAM \cite{papoudakis21liam}, MELIBA \cite{zintgraf21meliba} and ODITS \cite{gu2021online}. The implementation of CTCAT is based on deep recurrent Q-network \cite{hausknecht15drqn}, with the reward being adjusted by our proposed instance-wise teammate feedback rectification. The final behaviors of all baselines (including CTCAT) are determined by their most frequently picked choices from four candidate policies, with a consecutive $T$-step observation as input. By unifying the policy space of all baselines, the outcome is now solely determined by the robustness of different baselines against type confounding, whose level can be manually manipulated. As type confounding has not been discovered in previous works, no existing benchmarks can be directly used to quantitatively analyze its effect. Our experiment provides a unique platform that covers both necessary elements of ad hoc teamwork and tools for quantitatively analyzing type confounding. We now use experimental results to demonstrate the following observations: 
\begin{observation}\label{claim:1}
    Without rectification, the best response for a teammate type can be arbitrarily distorted by the distribution of teammate instances.
\end{observation}
            
This experiment adopts different teammate distributions to pre-train the agent's policy. Since the best response for all teammate instances in our experiment has been fixed to $\pi_1$, it is expected that modifying the distribution of teammate instances should not affect the choice of our agent, which is critical for its robustness against type confounding. The results are demonstrated in Table \ref{tab:overall_result}. Apart from the evenly distributed scenario, we modify the teammate distributions to let instances performing well on policy $\pi_2$, $\pi_3$ and $\pi_4$ have higher chances to be sampled, respectively. Theoretically, the agent can still obtain an unbiased outcome by comparing results of different policies for each instance. However, in practice, the agent's choice can be arbitrarily influenced by the setting of teammate distributions. This phenomenon is universal among our chosen baselines (including an ablation version of CTCAT that does not rectify the feedback). Instead, the performance of CTCAT is consistent among all scenarios, showing that it is more robust to type confounding. Therefore, the experimental result demonstrates our observation \ref{claim:1}, which supports the necessity of instance-wise teammate feedback rectification. 
            
\begin{observation}\label{claim:2}
    Collecting more data cannot directly address the problem of type confounding, since it does not influence the dependency relationships of ad hoc teamwork. 
\end{observation}

\begin{figure}[t]
    \vskip 0.1in
    \begin{center}
        \includegraphics[width=\linewidth]{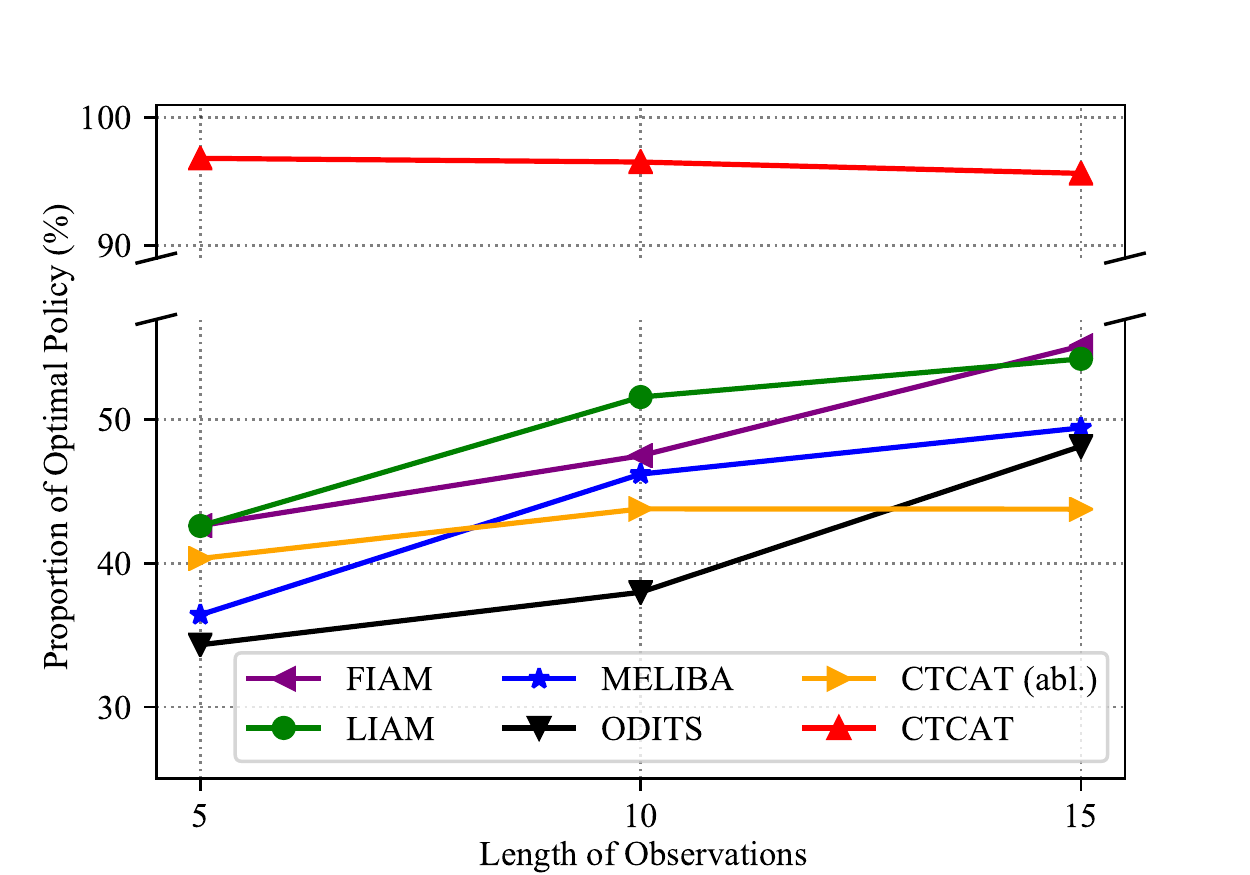}
        \caption{The impact of observation lengths on type confounding. The gaps among CTCAT and other baselines are significant. }
        \label{fig:timewindow}
    \end{center}
    \vskip -0.2in
\end{figure}

This experiment collects various lengths of observations as input to evaluate the influence of data size on type confounding. The distribution of teammate instances is set to spuriously favor the sub-optimal policy $\pi_2$. For each episode, we let the agent collect $T$ steps of observation before making the choice. During the data collection, the agent adopts a default no-op action to ensure the observation is not biased toward any candidate policy. In practice, letting the teammate operate without any feedback would affect the user experience. Therefore, the agent should adjust its policy as fast as possible. The experimental results under this setting are presented in Figure \ref{fig:timewindow}. We modify the length of observations to be 5, 10 and 15 steps, based on the fact that 
a pair of well-trained agents can finish the task with 15.86 steps  on average. Along with more teammate behaviors, the agent's proportions of picking the optimal policy have increased (about 10\%) for most comparing baselines, which is brought by more complete teammate information. However, their deviation with CTCAT is still huge (about 40\%). In this case, the impact of type confounding constitutes the major source of performance gap since very limited teammate behaviors can be further provided. 
Therefore, collecting more data cannot directly address the problem of type confounding, which supports our observation \ref{claim:2}.

\subsection{Real-world Scenarios}
\label{subsec:exp1}
            
\paragraph{Kidney stone treatment  \cite{simpsonKidneyStone_1986}. }  This scenario is derived from a real-life medical study, which compares two strategies of treating kidney stones. The summarized raw data are presented in Table \ref{tab:kidneystone}. Patients with large stones are more severe and have a slightly lower recovery rate than those with small stones. Nevertheless, for both conditions, open surgery achieves higher recovery rate. Therefore, classifying all patients into a same type of disease is reasonable since they have similar symptoms and share a common best response. However, when the data is aggregated, closed surgery becomes spuriously more favorable.  Based on this background, if an ad hoc teamwork agent is pre-trained with past treatment data and then successfully identifies an unknown teammate as the type of patent with kidney stones, which policy will the agent adopt?

\begin{table}[t]
    \caption{Kidney stone treatment (recovered / total patients). }
    \label{tab:kidneystone}
    \begin{center}
        \begin{tabular}{ccc}
            \toprule
            & Open surgery & Closed surgery \\ 
            \midrule 
            Small Stones                      & \phantom{00}81 / 87 = \textbf{0.93}   & 234 / 270 = 0.87   \\ 
            Large Stones                     & 192 / 263 = \textbf{0.73} &  \phantom{00}55 / 80 = 0.69   \\ 
            \midrule 
            Total                                  & 273 / 350 = 0.78 & 289 / 350 = \textbf{0.83} \\ 
            \midrule 
            UCB1                                & 0.14$_{\pm0.01}$               &   \textbf{0.83}$_{\pm0.01}$ \\
            EXP3                                 & 0.47$_{\pm0.01}$               &   \textbf{0.83}$_{\pm0.01}$ \\
            O. Q.                      & 0.82$_{\pm0.01}$               &   \textbf{0.84}$_{\pm0.01}$ \\ 
            T. S.        & 0.14$_{\pm0.01}$               &   \textbf{0.83}$_{\pm0.01}$ \\ 
            Q-Learning                       &  0.77$_{\pm0.01}$              &    \textbf{0.83}$_{\pm0.01}$                   \\ 
            CTCAT                             &  \textbf{0.84}$_{\pm0.02}$  &       0.79$_{\pm0.02}$                     \\ 
            \bottomrule
        \end{tabular}
    \end{center}
\end{table}

To verify the necessity of instance-wise teammate feedback rectification, we compare the result of CTCAT with standard Q-learning, UCB1 \cite{ucb1}, EXP3 \cite{exp3}, Optimistic Q (O. Q.) \cite{sutton2018reinforcement} and Thompson Sampling (T. S.) \cite{thompson1933likelihood}. Q-learning has been widely applied in many previous ad hoc solutions \cite{barrett17plastic,chen20aateam}. The other baselines are classic online learning solutions designed for similar context. We can observe that without rectification, standard Q-learning obtains a biased outcome, which is unfavorable for patients of both stone sizes. Due to type confounding, the causal relationship between the agent policy and the cooperation outcome is distorted, making the learned best response no longer reliable. In contrast, by rectifying the instance-wise teammate feedback, CTCAT successfully learns an unbiased result that is consistent with the conclusion of patients having any size of stones. 

\paragraph{Magazine renewal rate \cite{wagner1982simpson}. } In early 1979, the publishers of \textit{American History Illustrated} were pleased to find that their content in February had an overall renewal rate of 64\%, which was better than 51\% in January. Since the renewal rate was aggregated from several established subscription categories, the publishers examined each category to identify the major source contributing most to the rise of overall renewal rate. The results are shown in Table \ref{tab:pet_boarding_shop3}, which are counter-intuitive: of all five categories, the renewal rates in January are higher than those in February, which suggest readers prefer contents in January. Designing separate content for each category of readers is too costly to be implemented. Therefore, it is important to determine which content is more favorable by aggregating the commonality of readers from different sources. 

Compared with the former task, the teammate type in this scenario is composed of more detailed sub-groups, making the situation more complicated. However, this does not prevent the emergence of type confounding, leading to a phenomenon in which the aggregated conclusion disagrees with all its instances. In consequence, the robustness of standard Q-learning is affected, showing spuriously that the content of February is more favorable. Nevertheless, the conclusion of CTCAT is consistent with the result of every sub-group.  This indicates that our solution is robust to the impact of type confounding, even though the teammate type is formed with instances that are very diverse.

\begin{table}[t]
    \caption{Magazine renewal rate (renewed / total customers). }
    \label{tab:pet_boarding_shop3}
    \begin{small}
        \begin{center}
            \begin{tabular}{ccc}
                \toprule
                & January & February \\ 
                \midrule 
                Gift    & \phantom{00}2,918 / 3,594 = \textbf{0.81}   &  \phantom{0,0,}704 / 884 = 0.80  \\ 
                
                P. R.$\dagger$ &14,488 / 18,364 = \textbf{0.79}                      &  3,907 / 5,140 = 0.76 \\ 
                
                Direct Mail & \phantom{00}1,783 / 2,986 = \textbf{0.60}                &  1,134 / 2,224 = 0.51 \\ 
                
                S. S.$\ddagger$ & \phantom{0}4,343 / 20,862 = \textbf{0.21} & \phantom{0,0,}122 / 864 = 0.14  \\ 
                
                C. A.$\S$ & \phantom{0,000,0}13 / 149 = \textbf{0.09} & \phantom{0,0,000}2 / 45 = 0.04 \\ 
                
                \midrule 
                Total & 23,545 / 45,955 = 0.51 & 9,157 / 5,869 = \textbf{0.64} \\ 
                
                \midrule
                UCB1 & 0.11$_{\pm0.01}$ & \textbf{0.64}$_{\pm0.01}$ \\ 
                EXP3 & 0.27$_{\pm0.01}$ & \textbf{0.65}$_{\pm0.01}$ \\ 
                O. Q. & 0.63$_{\pm0.01}$ & \textbf{0.65}$_{\pm0.01}$ \\ 
                T. S. & 0.07$_{\pm0.01}$ & \textbf{0.65}$_{\pm0.01}$ \\                                                                       
                Q-Learning                       &        0.51$_{\pm0.01}$      &  \textbf{0.65}$_{\pm0.01}$                    \\ 
                CTCAT                          &       \textbf{0.54}$_{\pm0.02}$      &  0.50$_{\pm0.01}$               \\ 
                \bottomrule
            \end{tabular}\\
            {$\dagger$: previous renewal; $\ddagger$: subscription service; $\S$: catalog agent.  }
        \end{center}
    \end{small}
\end{table}        

\section{Conclusion}
\label{sec:conclusion}

This work presents CTCAT to control the influence of type confounding in ad hoc teamwork. We first unveil the existence of type confounding and demonstrate the cause behind this phenomenon, which is due to spurious correlations brought by arbitrary distribution of teammate instances. Then, we propose to control type confounding by aligning the distribution of optimal cooperation outcomes with an unbiased one. In this way, the spurious correlation between the agent's policy and the cooperation outcome is untangled. The performance of CTCAT is evaluated in several domains, including classic ad hoc teamwork tasks and real-world scenarios. With detailed analysis, the effectiveness of CTCAT on controlling type confounding is demonstrated. 

Our work is the first attempt to study the problem of type confounding in ad hoc teamwork and we would like to point out several future directions. First, to simplify the problem setting, the experiments are based on two-player ad hoc teamwork environments. As agents in the real world often interact in an open environment, it is worth investigating the influence of type confounding in tasks with more agents, and we believe the work of Rahman et al. (\citeyear{rahman21openaht}) provides a good direction. 
Second, our work is based on the premise of zero communication between players. However, it has been demonstrated that communication is helpful to identify the teammate's real intention in ad hoc teamwork \cite{mirsky20apenny}, which we believe is also helpful to untangle the confoundedness between them.
                    
\section*{Acknowledgements}

This work was supported by STI 2030 Major Projects (2021ZD0200400), Natural Science Foundation of China (U1909202, 61925603), The Key Research and Development Program of Zhejiang Province in China (2020C03004), Zhejiang Lab and China Scholarship Council. Bo An is supported by the National Research Foundation, Singapore under its Industry Alignment Fund – Pre-positioning (IAF-PP) Funding Initiative. Any opinions, findings and conclusions or recommendations expressed in this material are those of the author(s) and do not reflect the views of National Research Foundation, Singapore.


\bibliography{causalAH}
\bibliographystyle{icml2023}

\newpage
\appendix
\onecolumn
\section{The Derivation of CTCAT}\label{app:pseudo_code}

\begin{figure*}[h]
    \vskip 0.2in
    \begin{center}
        \includegraphics[width=0.8\linewidth]{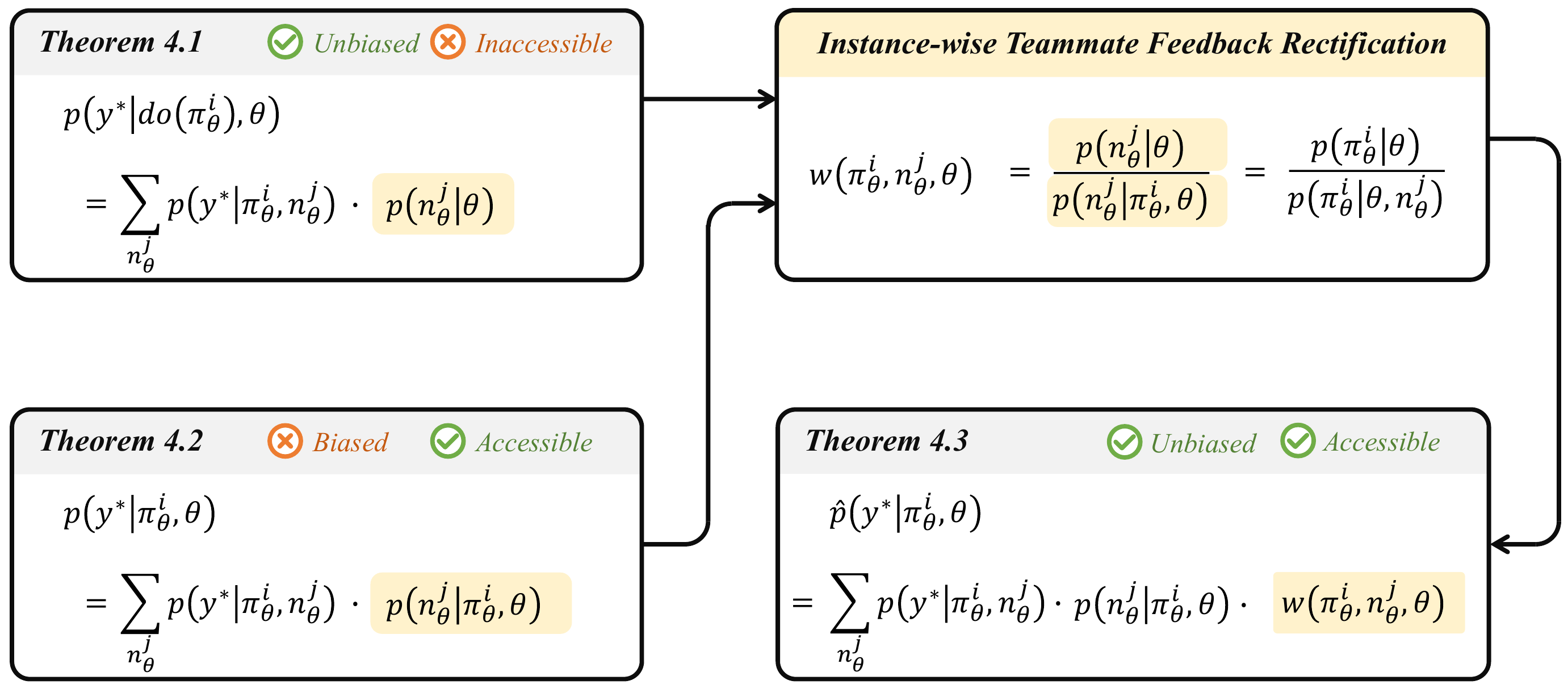}
        \caption{A diagram showing the relationships among Theorem \ref{prop:ideal}, \ref{prop:practical} and \ref{prop:combine}. }
        \label{fig:graph}
    \end{center}
\end{figure*}

The core of CTCAT is an instance-wise teammate feedback rectification method. Its theoretical background is to rectify the practical distribution of $y^\star$ (Theorem \ref{prop:practical}, biased but accessible) to make it align with the ideal distribution of $y^\star$ (Theorem \ref{prop:ideal}, unbiased but inaccessible). This procedure is achieved by our proposed instance-wise teammate feedback rectification, which is implemented by adopting $p(\pi_\theta^i \mid \theta, n_\theta^j)$ and $p(\pi_\theta^i \mid \theta)$ to rectify the feedback of each teammate instance (Theorem \ref{prop:combine}, unbiased and accessible). Figure \ref{fig:graph} depicts an overall diagram to demonstrate the relationships among Theorem \ref{prop:ideal}, \ref{prop:practical} and \ref{prop:combine} when deriving CTCAT. Algorithm \ref{alg:citycat} provides a pseudocode of CTCAT based on Q-learning.\footnote{Our algorithm can be naturally extended to other settings, such as PG-based \cite{yang21aaai} or TD-based solutions \cite{yang18ijcai}. }  
This implementation is adopted in Section \ref{subsec:exp1} where CTCAT is applied in two real-world scenarios. 

\begin{algorithm}[h]
    \caption{CTCAT}
    \label{alg:citycat}
    \begin{algorithmic}
        \STATE {\bfseries Input:} A stream of $( n_\theta^j, \pi_\theta^i, y )$ which represents the agent's collected interactions with teammates of type $\theta$
        \STATE Initialize the counter of $C( \pi_\theta^i \mid \theta )$ and $C( \pi_\theta^i \mid \theta, n_\theta^j  )$ for all $\pi_\theta^i$ and $n_\theta^j$ 
        \WHILE{not converged}
        \STATE  Update $C( \pi_\theta^i \mid \theta ) $ and $C( \pi_\theta^i \mid \theta, n_\theta^j )$ with respect to the sampled $( n_\theta^j, \pi_\theta^i, y )$
        \STATE Calculate the runtime estimations of $p( \pi_\theta^i \mid \theta )$ and $p( \pi_\theta^i \mid \theta, n_\theta^j )$ 
        \STATE Adjust $r$ by $\hat{r} = r \cdot p( \pi_\theta^i \mid \theta ) / p( \pi_\theta^i \mid \theta, n_\theta^j  )$ where $r$ is the reward of agent $i$ when interacting with teammate $n_\theta^j$
        \STATE Use $\hat{r}$ as the rectified reward to update  the agent's Q-values
        \ENDWHILE
    \end{algorithmic}
\end{algorithm}

\section{Experimental Setting}\label{app:experimental_setting}

\subsection{Goal-Based Predator-Prey}
\begin{figure}[t]
    \vskip 0.2in
    \begin{center}
        \includegraphics[width=0.27\linewidth]{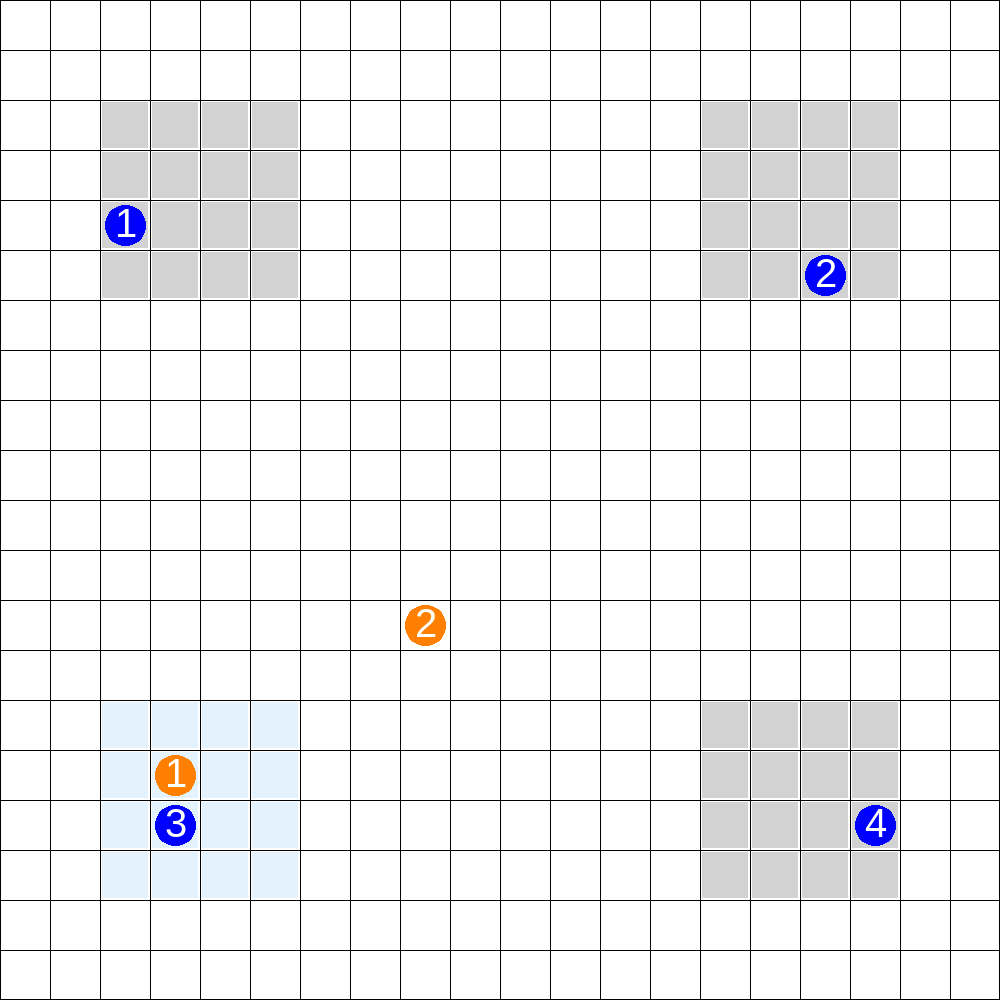}
        \caption{A screenshot of goal-based predator-prey where both predators choose the third prey as their goals. }
        \label{fig:game}
    \end{center}
    \vskip -0.2in
\end{figure}

\paragraph{Environmental Setting}
In our implementation of the goal-based predator-prey, the world is a toroidal grid map of size $20 \times 20$. If an agent moves off one end of the map, it appears on the other end. There are four preys on the map, marked as blue balls. The range of movement for each prey is limited to a $4 \times 4$ fenced area. Meanwhile, there are two predators, marked as orange balls, and one of them is under our control. Each predator has a private list of goals, which is a non-empty combination of the four preys. For each timestep, all agents (predators and preys) choose to either move into a neighboring cell or stay at their current position. If two agents run into the same cell, the collision is solved randomly. The prey's movement is determined randomly, but it will stay still if the prey tries to move out of the boundary or a predator is at its neighboring cells. To simulate the partial observation, each predator is limited to observing the coordinates of itself, its teammate, and the closest prey. The game terminates when both predators capture their preys or a maximal step of 300 has been reached. The prey is identified as being successfully captured if it is surrounded by two predators simultaneously. The predator receives a reward of $+1$ if it successfully captures a prey with its teammate and the prey is in its goal list. Otherwise, it receives a penalty of $-1$ when the game terminates. Figure \ref{fig:game} presents a screenshot of the goal-based predator-prey, where both predators choose the third prey as their common goal. 

\paragraph{Evaluation Setting}
To quantitatively evaluate the impact of type confounding, the predator's candidate strategies are trained in advance with self-played goal-based reinforcement learning \cite{ghosh2021goalbased}, and the goal of our controlled agent is to pick the most suitable strategy from the prepared  candidates. We let the agent collect interactions with 10,000 teammate instances using the previously prepared candidate strategies. All of these interactions are sampled from a fixed distribution of teammate instances, which 
enables us to control the level of type confounding by adjusting the teammate distribution. These interactions are stored in a replay buffer which is used for the agent's own policy training. The controlled agent's policy is determined by a consecutive $T$-step observation. For each timestep, the agent records its choice of candidate policy.  Nevertheless, within the initial $T$-steps,  the agent conducts a default no-op action so that the observation will not be influenced by its current decision. This prevents the observation from being biased toward any candidate policies. The agent's strategy after $T$-step is decided by the most frequently picked candidate policy. This setting ensures that the criteria of optimal policy for current teammate is common for all baselines, which makes quantitatively evaluating the impact of type confounding feasible since the major factor now affecting the agent's choice is 
its robustness against confoundedness. 

\paragraph{Teammate Instances and Candidate Policies}

The goal list of $K_1$ is $\{1, 4\}$, $K_2$ is  $\{1, 2, 3\}$, $U$ is  $\{1, 2, 4\}$. The goal lists for our four candidate policies ($\pi_1$ to $\pi_4$) are  $\{1, 2, 3\}$,  $\{2, 3\}$,  $\{2\}$ and  $\{3\}$, respectively. In this setting, the first prey appears on all $K_1$, $K_2$ and $U$, but is only selected in $\pi_1$ among our four candidate policies. Therefore,  $K_1$, $K_2$ and $U$ share similar behaviors when they choose to pursue the first goal, and  $\pi_1$ is the best response for all teammate instances in this case. Table \ref{tab:predator_prey} presents the success rate of our selected teammate instances ($K_1$, $K_2$ and $U$) and the candidate policies ($\pi_1$ to $\pi_4$), which are obtained with 10 independent runs. 

\begin{table}[h]
    \caption{Success rates between the sampled instances and the candidate strategies.  }
    \label{tab:predator_prey}
    \begin{center}
        \begin{tabular}{ccccc}
            \toprule
            & $\pi_1$ & $\pi_2$ & $\pi_3$  & $\pi_4$ \\ 
            \midrule 
            $K_1$ &  \textbf{0.40}$_{\pm0.07}$ & 0.05$_{\pm0.02}$  & 0.00$_{\pm0.01}$ & 0.01$_{\pm0.01}$  \\ 
            $K_2$ &  \textbf{0.96}$_{\pm0.02}$ & 0.88$_{\pm0.03}$  & 0.76$_{\pm0.04}$ & 0.57$_{\pm0.04}$  \\ 
            \midrule 
            $U$ &  \textbf{0.73}$_{\pm0.05}$ & 0.39$_{\pm0.03}$  & 0.64$_{\pm0.05}$ & 0.01$_{\pm0.01}$  \\ 
            \bottomrule
        \end{tabular}
    \end{center}
\end{table}

\paragraph{Distributions of Teammate Instances }

\begin{figure}[h]
    \vskip 0.2in
    \begin{center}
        \begin{minipage}{0.4\linewidth}
            \begin{center}
                \includegraphics[width=\linewidth]{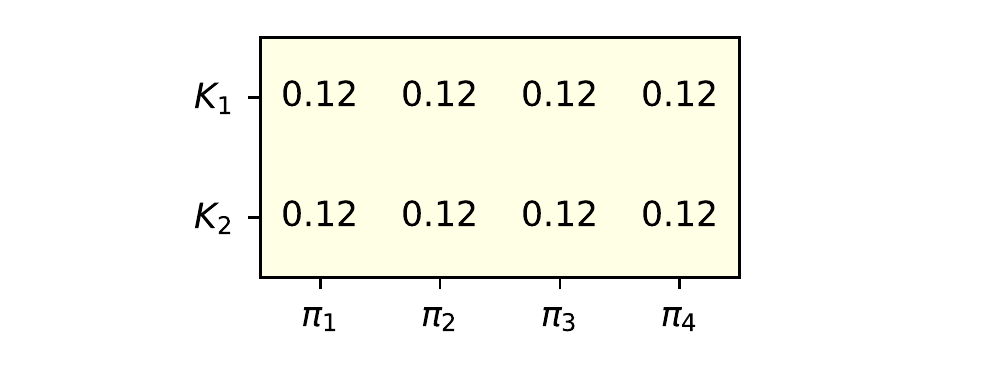}  \\ 
                (a)  The data are evenly distributed. 
            \end{center}
        \end{minipage}
        \begin{minipage}{0.4\linewidth}
            \begin{center}
                \includegraphics[width=\linewidth]{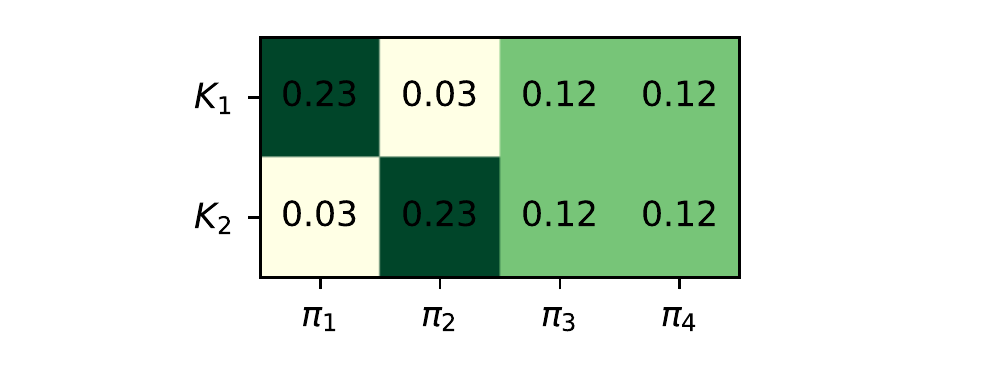}  \\ 
                (b)  $\pi_2$ is spuriously more favored. 
            \end{center}
        \end{minipage}        
        
        \begin{minipage}{0.4\linewidth}
            \begin{center}
                \includegraphics[width=\linewidth]{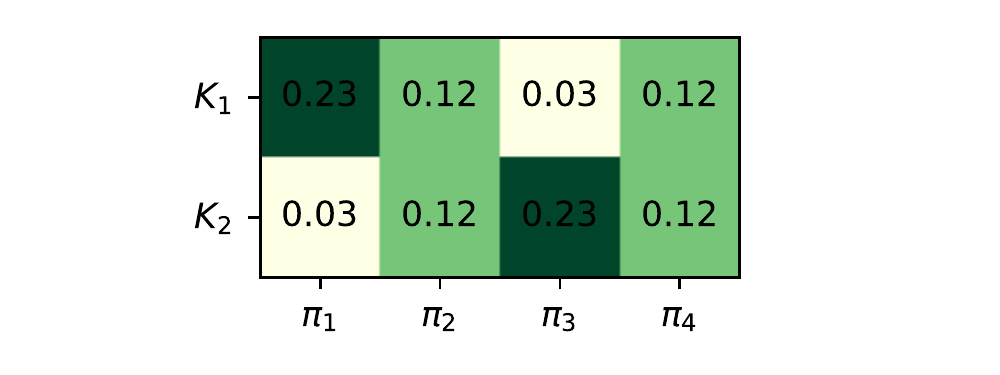}  \\ 
                (c)  $\pi_3$ is spuriously more favored. 
            \end{center}
        \end{minipage}
        \begin{minipage}{0.4\linewidth}
            \begin{center}
                \includegraphics[width=\linewidth]{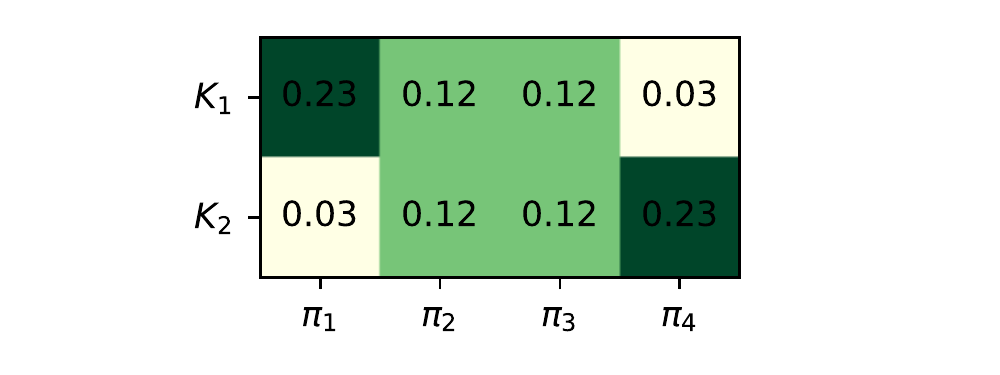}  \\ 
                (d)  $\pi_4$ is spuriously more favored. 
            \end{center}
        \end{minipage}         
        \caption{The sampled distributions of teammate instances which lead to type confounding. }
        \label{fig:sample_dist}
    \end{center}
    \vskip -0.2in
\end{figure}

Figure \ref{fig:sample_dist} presents the probabilities of each teammate instance being sampled 
when collecting replay buffers for the agent's pre-training. The standard used to distort the distribution of teammate instances is that interactions with sub-optimal outcomes have higher chance to be sampled. In many practical scenarios, the agent cannot determine the distribution of teammate instances and needs to learn from existing data. This makes the risk of type confounding emerge unnoticeably. 

\subsection{Kidney Stone Treatment \& Magazine Renewal Rate}

In these two tasks, the distributions of teammate instances are obtained from real-world scenarios where type confounding 
distorts the causal correlations between the agent's policy and the cooperation outcome. To obtain a statistically meaningful 
outcome, a random noise is added to the probability of each teammate instance receiving positive feedback.  This noise is sampled from a normal distribution $\mathcal{N}(0, 0.1)$. The baseline used for comparison in each task is implemented with a vanilla tabular Q-learning, and CTCAT is implemented by rectifying the reward of tabular Q-learning with our proposed instance-wise teammate feedback rectification. All the experiments are conducted with 10 independent runs to obtain the statistical outcomes. 

\section{Baselines} 

\subsection{LIAM \cite{papoudakis21liam}} 
LIAM adopts an encoder-decoder network to reconstruct the teammate's observations and actions for each time step with the agent's local observation history. The implementation of LIAM includes three components, which are the teammate encoder, the teammate decoder and the agent's policy network. The teammate encoder is implemented with a deep recurrent neural network. The teammate decoder and the agent's policy network are both implemented with feed-forward networks. For each timestep, the agent uses local observation as the encoder input to infer the teammate embedding. This embedding denotes the teammate's type from the agent's perspective. Then, the agent's policy network utilizes both the agent's local observation and the inferred teammate embedding as input to generate the corresponding action. During pre-training, the teammate encoder and decoder are simultaneously optimized to recover the teammate's local observation and chosen action. This setting encourages the agent to reconstruct the teammate's hidden representation with its local observation.

\subsection{FIAM \cite{papoudakis21liam}}
FIAM is an extension of LIAM.   The difference between them is that for FIAM, the agent's observation also includes the teammate's private observation. This enables FIAM to access more complete information than LIAM, even though this implementation is often considered unrealistic in many practical scenarios. 

\subsection{MELIBA \cite{zintgraf21meliba}} 
MELIBA includes three major components: teammate encoder, teammate decoder and the agent's policy network. The cascaded encoder-decoder network is used to infer the teammate's hidden embedding, which is similar to LIAM. However, the difference between them is that for MELIBA, the encoder-decoder is implemented with a variational auto-encoder (VAE) \cite{kingma13vae}. This enables the agent's learned distribution of teammate embedding to approximate an evidence lower bound of the optimal distribution which is defined in the context of VAE. 

\subsection{ODITS \cite{gu2021online}}  
ODITS includes six major components: the agent's proxy encoder/decoder, the policy network, the teamwork situation encoder/decoder and an integrated network. The proxy encoder-decoder and teamwork situation encoder-decoder form a pair of parallel pipelines to model the teammate type with either the agent's local observation or the global observation, respectively. The output of each encoder determines a Gaussian distribution, and the mutual information between these two distributions is minimized to let the agent correctly infer the teammate type with its private observation. The decoder output is used to determine the parameters of agent's policy network and the integrated network. In our implementation, the agent's policy network adopts both its own local observation and the output of proxy decoder as input to generate the expected value of each action. The parameters of integrated network are generated by the teamwork situation decoder.

\end{document}